\documentclass[twocolumn,showpacs,preprintnumbers,amsmath,amssymb]{revtex4}

\usepackage{graphicx}
\usepackage{dcolumn}
\usepackage{bm}

\begin{document}

\preprint{APS/123-QED}


\title{Electric-Field Breakdown of Absolute Negative Conductivity
and
Supersonic Streams 
in Two-Dimensional Electron Systems with Zero Resistance/Conductance
States 
} 
\author{ 
V.~Ryzhii}
\email{v-ryzhii@u-aizu.ac.jp}
\author{A.~Satou}
\affiliation{Computer Solid State Physics Laboratory, University of Aizu,
Aizu-Wakamatsu 965-8580, Japan}

\date{\today}

\begin{abstract}
We calculate the current-voltage characteristic 
of a two-dimensional electron system (2DES)
subjected to a magnetic field at strong electric fields.
The interaction of electrons with piezoelectric acoustic phonons is
considered as a major scattering mechanism governing the current-voltage
characteristic.
It is shown that at a sufficiently strong electric field corresponding
to the Hall drift velocity exceeding the velocity of sound,
the dissipative current exhibits an overshoot.
The overshoot of the dissipative current can result in a
breakdown of the absolute negative conductivity
 caused by microwave irradiation and, therefore,
substantially effect 
the formation of the domain structures with
the zero-resistance and zero-conductance
states and 
supersonic electron streams. 
\end{abstract}



\maketitle

Recently, two experimental groups~\cite{1,2}
have reported the observation of 
vanishing electrical resistance in two-dimensional
electron systems (2DES's) subjected to a magnetic field and
strong microwave radiation. 
The occurrence of the effect of the so-called zero-resistance (ZR)
and zero-conductance
(ZC) states~\cite{1,2,3}  
has been linked  to
the effect of  absolute negative conductivity (ANC)
when the dissipative dc conductivity $\sigma(E)$, determined by 
the dark conductivity $\sigma_{d}(E)$ and microwave
photoconductivity  
$\sigma_{ph}(E)$, is negative~\cite{4,5,6}.
As speculated, the mechanism of ANC  responsible
for the occurrence of ZR and ZC states is associated with
the photon-assisted scattering of electrons 
on impurities~\cite{7,8,9,10,11},
although alternative mechanisms, particularly the photon-assisted
interaction of electrons with acoustic phonons,
 can be essential as well~\cite{12,13,14,15}.
It has long been shown~\cite{7,8} that  the photon-assisted impurity
scattering can result in a negative dissipative photoconductivity $
\sigma_{ph}(E)$ 
in the situations in question 
when the  microwave frequency $\Omega$ somewhat exceeds
the cyclotron frequency $\Omega_c = eH/mc$ or its harmonics.
According to the calculations~\cite{7,8}, the dissipative microwave
photoconductivity
can be negative at sufficiently strong irradiation at
$0 < \Omega - \Lambda\Omega_c \lesssim eEL/\hbar, \Gamma/\hbar$.
Here $e$ and $m$ are the electron charge and effective mass, respectively, 
$c$ is the velocity of light,
$E$ and $H$ are the electric and
magnetic fields,
$L = (c\hbar/eH)^{1/2}$ is the quantum Larmor radius, 
$\Gamma$ is the Landau level (LL)
broadening, and $\hbar$ is the Planck constant. 

It is crucial for the explanations of ZR and ZC states, invoking 
the concept of ANC, in
a 2DES with both  the Hall bar and  Corbino 
configurations  that there is an electric field $E_0$
at which $\sigma(E_0) = 0$~\cite{5,6,16,17}.
The modulus of the dissipative photoconductivity
decreases with increasing electric field when the latter becomes
sufficiently large~\cite{7,8}, namely, when
 $E >  E_b =  max \{\hbar(\Omega - \Lambda\Omega_c)/eL,\, \Gamma/eL\}$. 
Hence, one can expect that 
$E_0$ is determined  by the resonance detuning and the LL broadening
with  $E_0 >  E_b$. The characteristic field
$E_b$ is rather large. Indeed, at the magnetic field $H = 2$~kG,
assuming that $|\Omega - \Omega_c|/\Omega_c = 0.25$,
we have $E_b \simeq 20$~V/cm. 
The estimated values of $E_b$ significantly exceed
the average electric field  $\langle E \rangle$
in the 2DES observed experimentally,
which  is  
in the range $\langle E \rangle \simeq 3\times(10^{-3} - 10^{-1})$~V/cm
depending on the sample geometry.
Taking into account the instability of uniform electric-field
distributions at $E < E_0$, one can  conclude 
(see, for example, Refs.~\cite{5,6,11}), that in the case
$\langle E \rangle \ll E_0$,
nearly periodic domain
structures are formed as shown in Fig.~1.
In these domain structures,
$|J_y| = \sigma_Hd|\langle E_x\rangle|  \ll \sigma_HdE_0$,\, $J_x = 0$,
and  $E_y =0$ in the Hall bar configuration, and
$|V_y| = d|\langle E_y\rangle| \ll dE_0$,\, $J_y = 0$, and $E_x = 0$
in the Corbino samples.
\begin{figure}
\begin{center}
\includegraphics[width=8.5cm]{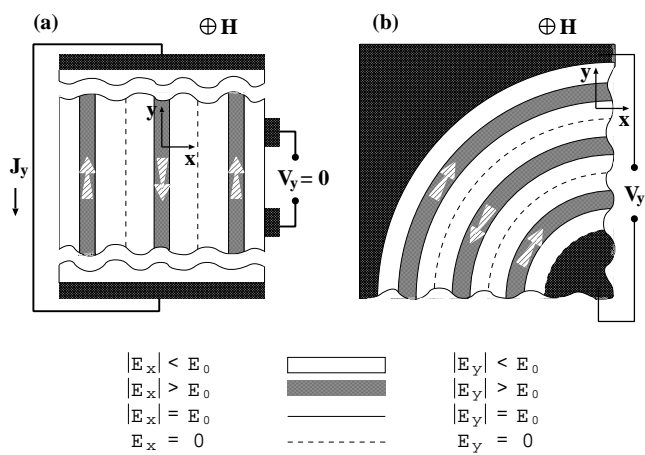}
\end{center}
\label{fig1}
\caption{Schematic view of possible domain structures  in 2DES 
corresponding to (a) ZR states 
in the  Hall bar configuration and input current
($J_y$ is the input current and $V_y = 0$ is the measured
voltage) and (b) ZC states  in the
Corbino geometry (here $V_y$ is the applied voltage).
Arrows show directions of the Hall current in different
high-field domains.
}
\end{figure}
\begin{figure}
\begin{center}
\includegraphics[width=7.0cm]{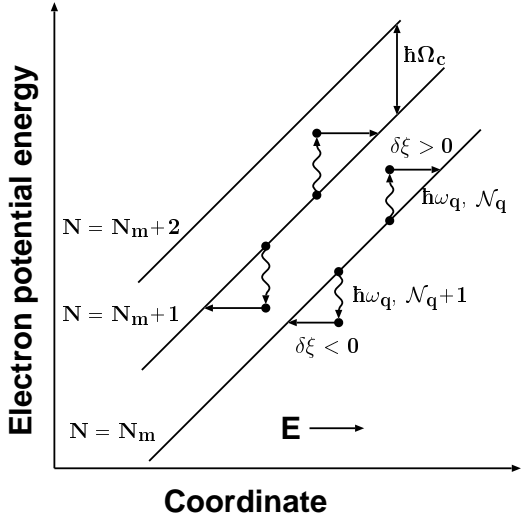}
\end{center}
\label{fig2}
\caption{Intra-LL electron transitions with absorption
and emission of acoustic phonons and spatial shift of electron  Larmor
centers  $\delta \xi$.
}
\end{figure}
Here $\sigma_H$ is the Hall conductivity and $d$ is the 2DES size.
It would appear reasonable that the shape of the domains and their
parameters,
in particular the swing of the electric field,   depend
on the behavior of the current-voltage characteristic, especially
at $E \simeq E_0$. Thus, mechanisms 
providing ANC at low electric field are essential for the instability
of uniform states that, as a matter of fact, result in the formation of the domains, 
whereas a mechanism
providing the lowest value of $E_0$ can 
substantially govern the domain structure and, therefore,
the observable macroscopic characteristics.

The electric-field
dependence of the net dissipative conductivity can be determined not
only by $\sigma_{ph}(E)$ but by $\sigma_{d}(E)$ as well. As shown
in the following, the dark component can be a steep
function of the electric field resulting in the breakdown of ANC
if $\sigma_{ph}(E) < 0$. 
The contribution of the
electron-impurity interaction to $\sigma_{d}(E)$
results in a smooth electric-field dependence up to very
strong electric fields $E > E_c = \hbar\Omega_c/eL$~\cite{18,19,20}
(see also Ref.~\cite{11}).
However, the contribution of the electron scattering on
acoustic phonons, being fairly small at low temperatures
and weak electric fields~\cite{13,14,21,23},
can lead to a sharp overshoot of the dissipative
current (conductivity) at  $E \simeq E_s = \hbar s/eL^2$, where $s$ is the
velocity of
sound. The point is that at 
$E < E_s$, the 
selection rules allow only the 
electron-phonon scattering events accompanied by
the inter-LL electron transitions. Since the energy of the acoustic
phonons involved equals the LL separation $\hbar\Omega_c$,
and therefore is rather large,
the number of such phonons at low temperatures is exponentially small.
This results in relatively small contribution
of the electron-phonon interaction to the
 dissipative dark current (dissipative
conductivity) at $E < E_s$. 
In contrast, at $E_s \lesssim E \ll E_c$, 
the intra-LL scattering  transitions 

can significantly contribute to the dissipative dark current.
Such transitions are schematically shown in Fig.~2.
This range of the  electric fields 
corresponds to  the electron Hall drift velocity 
$v_H = cE/H \geq s$.
Assuming $s = 3\times10^5$~cm/s and $H = 1 - 2$~kG, 
one can obtain 
$E_s \simeq 3 - 6$~V/cm. 
A peculiarity of the electron transport at $E \simeq E_s$
was pointed out previously, particularly
 in connection with the breakdown
of the quantum Hall effect~\cite{21,22,23,24,25,26,27} (see
also Refs. therein).

The dissipative  dark current density $j_d(E)$ and, consequently,
the dissipative dark conductivity $\sigma_{d}(E) = j_d/E$ 
of a
2DES in the case of the electron-phonon interactions
 can be calculated
using the following formula:
\begin{equation}\label{eq1}
j_{d}(E) = j_1 + j_2
= \sum_{N \neq N^{\prime}}j_{N,N^{\prime}} + \sum_{N}j_{N,N},
\end{equation}
where
$$
j_{N,N^{\prime}} = \frac{e}{\hbar} 
f_N(1 - f_{N^{\prime}})
$$
$$
\times\int d^3{\bf q}\,
q_y|V_{{\bf q}}|^2|Q_{N,N^{\prime}}(L^2q^2_{\perp}/2)|^2
$$
$$
\times\{
{\cal N}_q
\delta[(N - N^{\prime})\hbar\Omega_c + \hbar\omega_q + eEL^2q_y]
$$
\begin{equation}\label{eq2}
+ ({\cal N}_q + 1)
\delta[(N - N^{\prime})\hbar\Omega_c - \hbar\omega_q + eEL^2q_y] 
\}.
\end{equation}
Here
$f_N$ and  ${\cal N}_q$,
are the electron and phonon
distribution functions, 
$N = 0,1,2,...$ is the  LL index,
 ${\bf q} = (q_x, q_y, q_z)$, ${\bf q}_{\perp} = (q_x, q_y)$,
and $\omega_q = sq$
are 
the phonon wave vector, its in-plane component,  and the phonon frequency, 
respectively,
$\delta (\omega)$ is  the
LL form-factor determined by
$\Gamma$,  
$|V_{{\bf q}}|^2 = |C|^2\exp (- l^2q_z^2/2)/q $ 
describes  the electron  interaction with piezoelectric phonons,
where
 $l$ is
the width of the electron localization in the $z$-direction perpendicular to
the 2DES plane  and $|C|^2$ is a constant. 
At a strong localization in the quantum well at the heterointerface,
  $l \ll L$.   
$|Q_{N,N^{\prime}}(L^2q^2_{\perp}/2)|^2  = 
|P_N^{N^{\prime} - N}(L^2q^2_{\perp}/2)|^2 \exp (- L^2q^2_{\perp}/2))$ 
is determined by the overlap of the electron wave functions, and
$|P_N^{N^{\prime} - N}(L^2q^2_{\perp}/2)|^2$
is proportional to a Laguerre  polynomial 
$L_N^{N^{\prime} - N}(L^2q^2_{\perp}/2)$.
\begin{figure}
\begin{center}
\includegraphics[width=8.cm]{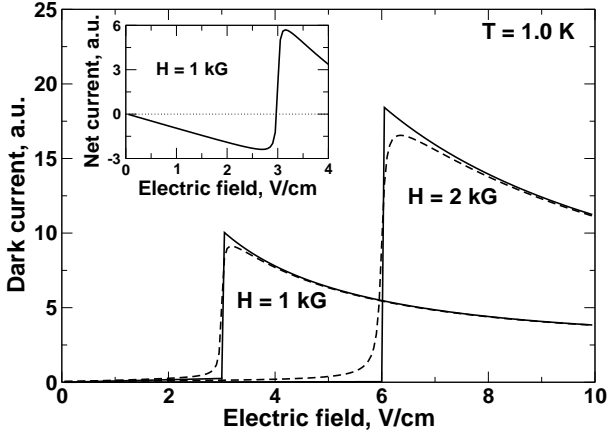}
\end{center}
\label{fig3}
\caption{Dissipative dark current-voltage characteristics
for different magnetic fields.
The dashed curve corresponds to the LL broadening $\Gamma/\hbar\Omega_c = 0.01$.
Inset shows net current-voltage characteristic
at microwave irradiation.}
\end{figure}

In the range of moderate electric
fields (much smaller than $E_c$),
the contribution of the electron scattering on acoustic phonons
accompanied by the inter-LL transitions 
can be presented as~\cite{12} 
\begin{equation}\label{eq3}
j_1 \simeq \sigma_1 E,
\end{equation} 
\begin{equation}\label{eq4}  
\sigma_1 = \mu_1 \biggl(\frac{\hbar s}{TL}\biggr) 
\biggl(\frac{s}{L\Omega_c}\biggr) \exp\biggl(- \frac{\hbar\Omega_c}{T}\biggr)
\exp\biggl(- \frac{l^2\Omega_c}{2s^2}\biggr),
\end{equation} 
where $\mu_1 \propto |C|^2\sum_Nf_N(1 - f_{N + 1})/\sqrt{N}$ and 
$T$ is the temperature (in energy units). The quantity
$j_1$ remains  small up to very large electric 
fields~\cite{23} (as in the case of elastic
impurity scattering~\cite{18,20}) $E \simeq E_c$.

As for $j_2$  corresponding to the  electron scattering 
on acoustic phonons
accompanied by the intra-LL transitions, 
it equals zero at $ E  <  E_s = \hbar s/eL^2$ because 
the scattering selection rule $\hbar \omega_q = eEL^2q_y$
is not met  at low electric fields. 
At $E \geq E_s$, considering Eqs.~(1) and (2), 
we obtain
$$
j_2 = \frac{e|C|^2}{\hbar^2s} 
\sum_Nf_N(1 - f_{N})\int d\Phi \sin\Phi dq\,dq_{\perp}
\frac{q^2_{\perp}}{\sqrt{q^2 - q^2_{\perp}}}
$$
$$
\times
\exp(- l^2q^2/2)\exp[(l^2- L^2)q^2_{\perp}/2]
|L_{N}^{0}(L^2q^2_{\perp}/2)|^2
$$
\begin{equation}\label{eq5}
\times[
{\cal N}_q
\delta(q + Fq_{\perp}\sin\Phi)+ ({\cal N}_q + 1)
\delta(q - Fq_{\perp}\sin\Phi)].
\end{equation}
Here $F = eEL^2/\hbar s = cE/sH$ and  $\sin\Phi = q_y/q_{\perp}$. 
In contrast to Ref.~\cite{25}, 
we will focus on
the case of large filling factors corresponding  to the
experimental conditions~\cite{1,2,3}, i.e., the case $\zeta \gg
\hbar\Omega_c$, where $\zeta$ is the Fermi energy
reckoned from the lowest LL. In this case, the LL's with
large $N$ yield the main contribution (with $N = N_m$ and $N = N_m + 1$,
where $\zeta/\hbar\Omega_c  \leq  N_m < \zeta/\hbar\Omega_c + 1$). 
When $N \gg 1$, $|L_{N}^{0}(L^2q^2_{\perp}/2)|^2 
\simeq J_0^2(\sqrt{2N}Lq_{\perp}) 
\simeq 2
\cos^2(\sqrt{2N}Lq_{\perp} -\pi/4)/\pi\sqrt{2N}Lq_{\perp}$, 
where $J_0 (z)$ is the Bessel function.
Taking this into account, after integrating  over $q$
Eq.~(5) can be reduced to 

\begin{figure}
\begin{center}
\includegraphics[width=8cm]{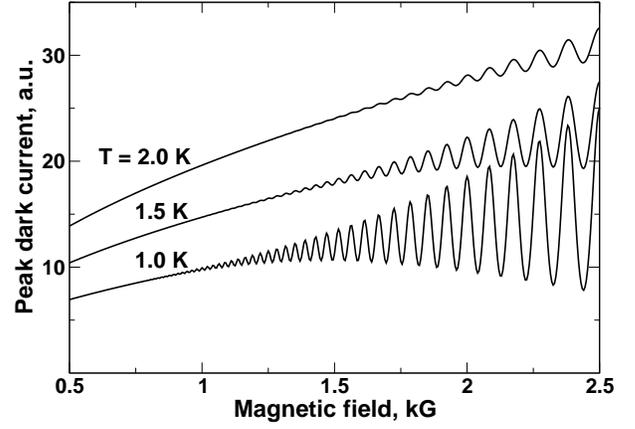}
\end{center}
\label{fig4}
\caption{Peak value of dark current vs
magnetic field at different temperatures ($20 \leq \zeta/\hbar\Omega_c 
\leq 100$).
}
\end{figure}
$$
j_2 =
\frac{\mu_2\hbar s}{2eL}\int d\Phi\sin\Phi dq_{\perp}
\frac{\exp[- l^2(F^2\sin^2\Phi - 1)q_{\perp}^2/2]}
{\sqrt{F^2\sin^2\Phi - 1}}
$$
\begin{equation}\label{eq6}
\times\exp(-  L^2q_{\perp}^2/2)\cos^2[\sqrt{2N}Lq_{\perp} - \pi/4 ].
\end{equation}
Here
$\mu_2 \propto |C|^2\sum_{N}f_N(1 - f_{N})/\sqrt{N}$.
One can see that the phonon distribution function
is excluded from the expression for the dissipative current.
This is because the contributions
of the processes of the stimulated emission of phonons
and their absorption compensate each other in the intra-LL transitions
(see Fig.~2).
Further integration in the right-hand side of Eq.~(6) over $q_{\perp}$
and $\Phi$ yields
$$
j_2 
= \frac{\sqrt{\pi}\mu_2\hbar s}{2\sqrt{2}eL^2}  
\int_0^{\sqrt{F^2 - 1}/F} \frac{d\cos\Phi} 
{\sqrt{(F^2- 1) - F^2\cos^2\Phi}}
$$
\begin{equation}\label{eq7}
\times\frac{1}{\sqrt{1   + a^2(F^2 - 1) -
a^2 F^2\cos^2\Phi }}
\simeq \frac{\pi^{3/2}\mu_2\hbar s}{2^{5/2}eL^2}
\frac{1}{F}
\end{equation}
with $a = l/L \ll 1$.

Taking into account Eqs.~(1), (3),  and (7),
we arrive at the following current-voltage characteristic:
\begin{equation}\label{eq8}
j_{d}(E) =  \sigma_1 E + \frac{\pi^{3/2}\mu_2}{2^{5/2}}
\frac{E_s^2}{E}\Theta(E - E_s),
\end{equation}
where $\Theta(E)$ is the unity-step function, which in the case of marked
LL broadening should be replaced by a smoother function.
Figure~3 shows the dark current-voltage characteristics calculated
using the above
formulas. It exhibits a pronounced overshoot of the dark current. 
The electric-field dependence of the dissipative conductivity
(with a jump at $E \simeq E_s$)
similar to that following from Eq.~(8) was recalculated from
the experimental data  at rather strong magnetic field~\cite{28}
and, consequently, proportionally large $E_s$ (see also Ref.~\cite{29}). 

As follows from the obtained expressions, the dissipative current
at $E > E_s$ depends on the temperature only via the factor $\mu_2$.
This factor provides a nontrivial dependence
of the dissipative conductivity, particularly at its peak magnitude, on the magnetic
field. Figure~4 shows the peak value (at $E = E_s$) 
of the dissipative current as a function of the magnetic field.
The pronounced oscillations  in Fig.~4 are akin to the Shubnikov-de Haas 
oscillations.

The relative height of the current overshoot is given by

\begin{equation}\label{eq9}
\frac{j_2}{j_1}\biggr|_{E= E_s} \simeq 
\biggl(\frac{\mu_2}{\mu_1}\biggr)\biggl(\frac{T}{ms^2}\biggr)
\exp\biggl(\frac{\hbar\Omega_c}{T}\biggr) \gg 1.
\end{equation}
The ratio $\mu_2/\mu_1$ varies with the magnetic field
 because the LL filling factor depends 
on the LL positions with respect to the Fermi level. 
However, when $T < \hbar\Omega_c$,
the product  $(\mu_2/\mu_1)\exp(\hbar\Omega_c/T) \gg 1$
at all values of $(\zeta - N_m\hbar\Omega_c)$.
The parameter $T/ms^2$ exceeds unity  at 
$T > 50$~mK.

A sharp increase in the dissipative current at $E = E_s$
can  also occur due to the electron interaction with interface acoustic 
phonons. In this case,
owing  to specific electron-phonon scattering selection rules~\cite{30},
the dissipative current can reveal even sharper overshoot at $E = E_s$.

Because of a strong overshoot of the dissipative dark current
at $E = E_s$, the net dissipative conductivity of a 2DES
irradiated with microwaves can change its sign at $E_0 \simeq E_s$.
The net current voltage characteristic of a 2DES under
microwave irradiation with $\sigma_{ph} < 0$
(associated, for example, with the photon-assisted scattering on impurities)
is shown schematically in the inset in Fig.~3.
The latter characteristic 
corresponds to  an abrupt change in the sign of the
net dissipative conductivity, i.e., the breakdown of ANC,
at $E_0 \simeq 3$~V/cm.
A sharp transition from $\sigma(E) < 0$ at $E < E_0 \simeq E_s$
to   $\sigma(E) > 0$  when  $E$ exceeds $E_s$ can significantly influence
the domain structures resulting in  ZR and ZC states.
The remarkable feature of these domain structures is that 
the regions, where $E_x > E_s$ in the Hall bar configuration
or $E_y > E_s$ in the Corbino samples, shown in Fig.~1 
by the shaded ``lanes'' with arrows, are sources of the acoustic phonons
generated by the electron streams with supersonic Hall drift velocities.
The generation of acoustic phonons by these streams can lead to 
a pronounced deviation of the phonon system from equilibrium.
This, in turn, can affect the values and sign of
the dissipative conductivity~\cite{15}.
 
Worth noting the feasibility of  formation of  stable 
ZR states at relatively large currents.
These states can correspond to $\langle E_x \rangle \simeq \pm E_0$
when the electric-field distributions 
are  nearly uniform. The states in question can be
formed when $J_y \simeq \pm J_0
= \pm \sigma_HdE_0$. Taking into account that $\sigma_H = ec\Sigma/H$,
where $\Sigma$ is the electron sheet concentration, and assuming
$E_0 = E_s$, one can obtain 
$J_0 \simeq  esd\Sigma$. The quantity $J_0$ is by two orders
of magnitude larger than the currents in ZR states
observed experimentally. The states
with smooth   electric-field distributions
can arise in the Corbino samples at  the applied voltage $V_y \simeq
\pm V_0$
with
$V_0 = dE_0 \simeq \pm dE_s$. If $d = 0.25$~cm,  at $H = 1$~kG one obtains
$V_0 \simeq 0.75$~V. Due to the absence of dissipation, the Joule heating
can be disregarded despite relatively large values of the current
(applied voltage).

In summary, we calculated the electric-field dependence of
the dissipative dark current
associated with the scattering of electrons on acoustic phonons
in a 2DES subjected to a magnetic field.
The obtained current-voltage characteristic exhibits a strong overshoot
at a certain electric field. This field corresponds to the
Hall drift velocity equal to the speed of sound.
The overshoot of 
the  dark component of the
dissipative current can suppress the negative microwave
photoconductivity (breakdown of ANC)
in the regions with high electric field,
where supersonic electron streams occur,
and affect the formation 
of domain structures with the zeroth resistance/conductance.

The authors thank   
V.~Volkov and V.~Vyurkov
for fruitful discussions and  comments on the manuscript.

\end{document}